# Bistability, multistability and nonreciprocal light propagation in Thue-Morse multilayered structures

**Victor Grigoriev and Fabio Biancalana**

Nonlinear Photonic Nanostructures Research Group,
Max Planck Institute for the Science of Light,
Guenther-Scharowsky-Str. 1 / Bau 26 (R440),
Erlangen 91058, Germany

E-mail: victor.grigoriev@mpl.mpg.de

**Abstract**. The nonlinear properties of quasiperiodic photonic crystals based on the Thue-Morse sequence are investigated. The intrinsic spatial asymmetry of these one-dimensional structures for odd generation numbers results in bistability thresholds which are sensitive to the propagation direction. Along with resonances of perfect transmission, this feature allows to achieve strongly non-reciprocal propagation and to create an all-optical diode. The salient qualitative features of such optical diode action is readily explained through a simple coupled resonator model. The efficiency of a passive scheme, which does not necessitate of an additional short pump signal, is compared to an active scheme, where such a signal is required.

## 1. Introduction

The experimental discovery of quasicrystals [1] has given rise to a sudden breakthrough in the area of solid state physics. It had been assumed for a long time that a periodic arrangement of unit cells was a main prerequisite for the specific properties of crystalline matter. The structure of quasicrystals does exhibit long-range order or correlations between distant parts, but at the same time there is no underlying periodicity, in the sense that a shifted copy of the crystal never matches exactly the original one. In fact, quasicrystals represent an intermediate stage between random media and periodic crystals, effectively combining both localization properties as a result of short-range disorder and the presence of band gaps due to long-range correlations [2].

The concept of quasiperiodicity was readily transferred to photonic crystals and proved to be of great value for many practical purposes [3-6]. Photonic quasicrystals are deterministically generated dielectric structures with a non-periodic modulation of the refractive index. In the one-dimensional (1D) case, they can be formed by stacking together dielectric layers of several different types according to the substitutional sequence under investigation (Cantor, Fibonacci, Rudin-Shapiro, Thue-Morse, etc.) [7]. The Fibonacci sequence is of particular importance, since it leads to the existence of two incommensurable periods in the spatial spectrum of the structure. Such behavior is typical of sequences with a so-called pure point spectrum, which makes the Fibonacci sequence truly quasiperiodic, as a consequence of the appearance of Bragg-like peaks in the spatial spectrum [8]. This property has been demonstrated to be very valuable for nonlinear optical applications, such as third harmonic generation [9]. In fact, the latter is a two-stage process, and for each stage different phase matching conditions should be met. The Fibonacci sequence allows to fulfill both of them simultaneously and on the same crystal [10].

In contrast to the Fibonacci sequence, the Thue-Morse (ThM) sequence possesses a singular continuous spectrum, which is neither continuous nor singular [11-12]. Thus, strictly speaking, the ThM sequence is not quasiperiodic, but rather deterministically aperiodic. To an even different class belongs the Rudin-Shapiro sequence, which shows evidence of a continuous spectrum, analogous to the one exhibited by random sequences [13]. With a slight abuse of terminology, we shall always refer to the above sequences as quasiperiodic, and we shall focus our attention specifically to the ThM sequence, for reasons explained below.

The bandgaps of quasiperiodic multilayered structures, also called 'pseudo band gaps' [14] or 'fractal bandgaps' in the literature [15], often contain resonant states, which can be considered as a manifestation of numerous defects distributed along the structure. Their quality factors are not very large when compared with those of symmetrically placed defects inside conventional Bragg structures, but the mode profiles associated to these resonances are extended in space and, as a consequence, very suitable for the enhancement of nonlinear effects throughout the whole body of the crystal. In order to investigate the specific properties of bistability and multistability in quasicrystals, in the following we shall restrict our attention exclusively to 1D structures. As a representative case, we focus our attention to the ThM aperiodic multilayer, and in making this choice we were guided by the following two reasons. Firstly, the spatial asymmetry of the ThM sequence (of odd generation numbers) interplays with the nonlinearity and makes the transmission sensitive to the propagation direction [16]. Secondly, almost all resonances in ThM quasicrystals are resonances of complete (i.e. 100%) transmission, with some of them being situated well inside the pseudo band gap regions. Taken together, the above two properties provide favorable conditions for the design of an all-optical diode, i.e. a device that shows a strong contrast in the transmission between forward and backward incidence. A proper understanding of bistability phenomena in quasiperiodic crystals is very important for the understanding of the optical diode action described above.

The paper is organized as follows. In section 2, we briefly review the well-known linear properties of ThM quasicrystals and classify their resonances by using the method of trace maps. It is shown that the field profiles at resonance frequencies follow the pattern of Thue-Morse sequence and the classification suggested can be used as a measure of this self-similarity. In section 3, we show that the spatial asymmetry inherent to ThM quasicrystals of odd generation numbers can interplay with Kerr-nonlinearity in such a way that transmission become sensitive to the propagation direction. It is very important that the dynamics of this interplay can have a sudden jump due to bistability when the intensity of the incident light changes, and the thresholds for such jumps are different for the incidence from the left and from the right of the multilayer. In section 4, we apply a phenomenological model based on a coupled resonator model to explain this behavior analytically for both bistable and multistable cases. The general formulas describing nonlinear transmission spectra are derived and their relationship to the level of self-similarity in the field profiles is emphasized. In section 5, we propose a design of nonlinear optical diode based on the differences in bistability thresholds between forward and backward propagation. The efficiency of two schemes is compared: passive, when only one pulse is used, and active, when an additional short pump signal is applied to facilitate switching between stable branches of hysteresis. The last section summarizes the results and presents the conclusions. The appendix provides some details about the FDTD method used in the paper for the dynamical simulations of multilayered structures with instantaneous Kerr-nonlinearity.

## 2. Thue-Morse multilayers

There are two main approaches to generate 1D quasiperiodic sequences. The first makes use of a projection from a higher-dimensional space, while the second employs the so-called substitutional sequences [17]. Being strictly quasiperiodic, the Fibonacci sequence can be obtained by using both the above methods, and the presence of two incommensurable periods in their spatial spectra is implicitly related to the projection of a two-dimensional grid onto a one-dimensional line [18]. However, these two approaches are not equivalent to each other, and the use of substitutional sequences tends to be a more general procedure [19]. In particular, this is applicable to ThM sequences and makes it distinct from strictly quasiperiodic sequences. Sometimes, it is called a deterministic aperiodic sequence, in order to emphasize that it has more disorder than quasiperiodic ones and stands closer to random sequences [20]. On the other hand, the development of resonances when changing generation number in ThM quasicrystals resembles the analogous development in periodic structures to a certain extent and brings these two kinds of crystal closer to each other [21].

Two dielectric layers of different materials are required to compose ThM quasicrystals, which shall be indicated with the letters 'A' and 'B'. They should be arranged in the same way as in literal ThM sequence, which is governed by the following substitutional ('inflation') rules: $'A' \to 'AB'$, $'B' \to 'BA'$. Thus, starting from a single layer $S_0 = 'A'$, which is defined to be the ThM quasicrystal of 0th generation or $\text{ThM}_0$, one obtains $S_1 = 'AB'$, $S_2 = 'ABBA'$, $S_3 = 'ABBABAAB'$ and so forth, with each step giving a sequence of generation number increased by one (figure 1a). It can be shown that an additional recurrence relation follows from this definition, which holds for ThM sequences as single blocks: $S_{n+1} = S_n \tilde{S}_n$, where $n$ is the generation number. In this notation $\tilde{S}_n$ indicates a sequence 'conjugated' to $S_n$, where all letters are interchanged to their opposite as in the rule $'A' \leftrightarrow 'B'$.

It is interesting to note that there is only a slight difference between the ThM and Bragg (or periodic) sequences as far as inflation rules are concerned. More specifically, Bragg sequences possess the inflation rule $'A' \to 'AB'$, $'B' \to 'AB'$. Therefore, the growth rate of these sequences is the same, and the total number of layers is subject to a rapid (exponential) increase as $2^n$. The occurrence of layers 'A' and 'B' is also the same for both ThM and Bragg structures for any generation number. Nevertheless, such a small modification of the inflation rules leads to completely distinct transmission and localization properties.

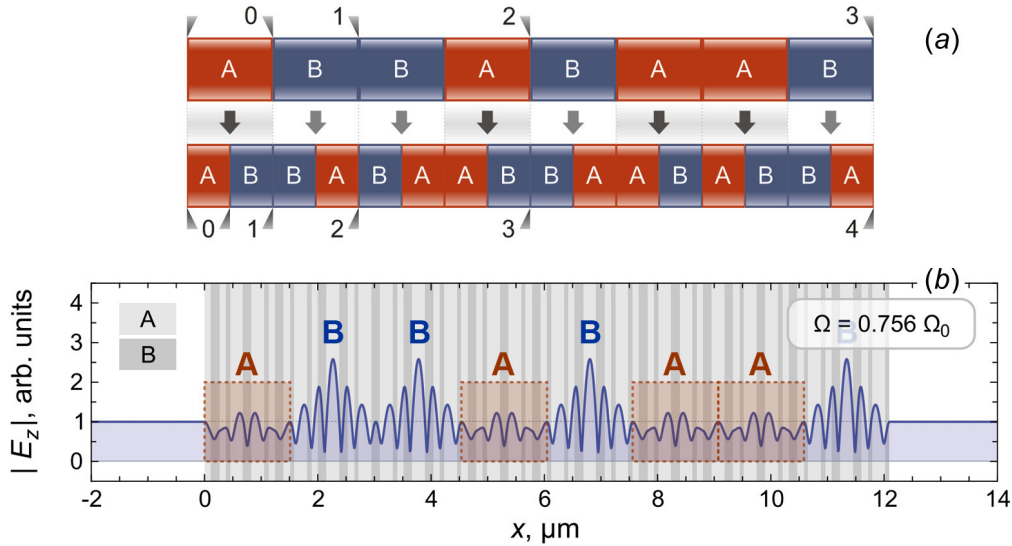

**Figure 1.** (*a*) Application of inflation rules $'A' \to 'AB'$, $'B' \to 'BA'$ for ThM sequence. Numbers near the notches designate the boundaries of sequences with corresponding generation numbers. (*b*) Field profile at one of the resonance frequencies ($\Omega/\Omega_0 = 0.756$) of a $\text{ThM}_7$ quasicrystal showing self-similar pattern of $\text{ThM}_3$ sequence. The frequency $\Omega_0$ corresponds to quarter wavelength condition taken at 0.7 μm. The linear refractive indices of the materials used are $n_A = 1.55$ (polydiacetylene 9-BCMU) and $n_B = 2.3$ (rutile). Background shows alternation of layers inside the structure.

The primary method to compute transmission spectra and field profiles for multilayered structures is the transfer matrix method [22]. If normal incidence is assumed, the electric $E$ and magnetic $H$ fields at the opposite sides of each single layer can be related in the following way:

$$\begin{pmatrix} E \\ iH \end{pmatrix}_{m-1} = \mathbf{M}_m \begin{pmatrix} E \\ iH \end{pmatrix}_m, \tag{1}$$

$$\mathbf{M}_m = \begin{bmatrix} \cos(n_m d_m \Omega) & n_m^{-1} \sin(n_m d_m \Omega) \\ -n_m \sin(n_m d_m \Omega) & \cos(n_m d_m \Omega) \end{bmatrix}, \tag{2}$$

where $d_m$ is the thickness of the layer with refractive index $n_m$, and $\Omega = \omega/c$ is the wavenumber in vacuum. In the representation of equation (1) the imaginary unit is deliberately kept with the magnetic field, in order to show explicitly that transfer matrices of nonabsorbing layered structures can be characterized by real matrix elements. In general, the refractive indices of the materials can depend on frequency, but for the sake of simplicity we will assume that all materials are dispersionless. By multiplying together all transfer matrices of subsequent layers, it is possible to construct the total transfer matrix of the structure

$$\mathbf{M} = \mathbf{M}_1 \mathbf{M}_2 \ldots \mathbf{M}_{L-1} \mathbf{M}_L, \tag{3}$$

and to find the transmission spectrum in terms of its matrix elements

$$T(\Omega) = \frac{4}{|M_{11} + M_{22} + i(M_{21} - M_{12})|^2}. \tag{4}$$

Another – more efficient – approach to obtain transfer matrices of ThM quasicrystals is to make use of the following two-step recurrence relations

$$\mathbf{M}_{\{n+1\}} = \mathbf{M}_{\{n\}} \tilde{\mathbf{M}}_{\{n\}}, \tag{5}$$

$$\tilde{\mathbf{M}}_{\{n+1\}} = \tilde{\mathbf{M}}_{\{n\}} \mathbf{M}_{\{n\}}, \tag{6}$$

supported by the initial conditions $\mathbf{M}_{\{0\}} = \mathbf{M}_A$ and $\tilde{\mathbf{M}}_{\{0\}} = \mathbf{M}_B$, which are the transfer matrices for layers 'A' and 'B', respectively. The subscripts in curly brackets indicate generation numbers of corresponding ThM quasicrystals. Formulas (5)-(6) take their origin in the recurrence relations of ThM sequence explained above, and stress explicitly the interplay between 'ordinary' $S_n$ and 'conjugated' $\tilde{S}_n$ counterparts of the sequence. The self-similarity of ThM quasicrystals is naturally embedded in this approach, and this simplifies considerably the computation of transmission spectra for large generation numbers.

However, even this modified representation of the transfer matrix method still contains redundant information about the structure, if one is interested only in the analysis of the properties of the resonance frequencies. In this case, it is sufficient to operate with traces of transfer matrices. The trace is defined as

$$x = \text{Tr}(\mathbf{M}) = M_{11} + M_{22}. \tag{7}$$

For ThM quasicrystals of fixed generation numbers, these traces are unchanged under conjugation

$$x_n = \text{Tr}(\mathbf{M}_{\{n\}}) = \text{Tr}(\mathbf{M}_{\{n-1\}} \tilde{\mathbf{M}}_{\{n-1\}}) = \text{Tr}(\tilde{\mathbf{M}}_{\{n-1\}} \mathbf{M}_{\{n-1\}}) = \text{Tr}(\tilde{\mathbf{M}}_{\{n\}}), \tag{8}$$

and for adjacent generation numbers, an additional relation or 'trace map' holds [23]:

$$x_{n+2} = \text{Tr}(\mathbf{M}_{\{n\}} \tilde{\mathbf{M}}_{\{n\}} \tilde{\mathbf{M}}_{\{n\}} \mathbf{M}_{\{n\}}) = \text{Tr}(\mathbf{M}_{\{n\}}^2 \tilde{\mathbf{M}}_{\{n\}}^2) = x_n^2 (x_{n+1} - 2) + 2. \tag{9}$$

The formulas (8)-(9) are valid for $n \geq 1$, and in derivation of (9) the Cayley-Hamilton theorem was applied to prove that any unimodular matrix satisfies the relation

$$\mathbf{U}^2 = x_U \mathbf{U} - \mathbf{I}, \qquad (10)$$

where $x_U$ is the trace of a unimodular matrix $\mathbf{U}$, and $\mathbf{I}$ is the identity matrix. To use the equation for the trace map (7), two initial conditions are required. They can be found directly from the multiplication of transfer matrices having the form given by equation (2):

$$x_1 = 2\cos\alpha\cos\beta - (n_A/n_B + n_B/n_A)\sin\alpha\sin\beta, \qquad (11)$$

$$x_2 = 2\cos 2\alpha\cos 2\beta - (n_A/n_B + n_B/n_A)\sin 2\alpha\sin 2\beta, \qquad (12)$$

where $\alpha = n_A d_A \Omega$ and $\beta = n_B d_B \Omega$.

Several important conclusions can be extracted from the trace map of ThM quasicrystals. The analysis simplifies greatly in absence of absorption, so that the imaginary parts of the refractive indices are negligible. In this case, all traces will be real, and there will be frequencies in the spectrum in correspondence to which $x_n = 0$. Independent of the value of $x_{n+1}$, this makes $x_{n+2} = 2$, which coincides with the trace of identity matrix. This simply means that frequencies for which $x_n = 0$ correspond to frequencies of perfect transmission. To prove this, an auxiliary matrix should be constructed, the trace of which gives the difference of diagonal elements of the original transfer matrix:

$$z = \mathrm{Tr}(\boldsymbol{\sigma}_z \mathbf{M}) = M_{11} - M_{22}, \qquad (13)$$

where $\boldsymbol{\sigma}_z$ is the third (diagonal) Pauli matrix. Provided that $x_n = 0$, it can be readily shown that the diagonal elements of $\mathbf{M}_{\{n+2\}}$ are equal:

$$z_{n+2} = \mathrm{Tr}(\boldsymbol{\sigma}_z \mathbf{M}_{\{n\}} \tilde{\mathbf{M}}_{\{n\}} \tilde{\mathbf{M}}_{\{n\}} \mathbf{M}_{\{n\}}) = \mathrm{Tr}[\boldsymbol{\sigma}_z] = 0. \qquad (14)$$

Therefore, the resonances of complete transmission in ThM quasicrystals can be characterized by identity transfer matrices, and they are preserved from one generation to another. Only additional resonances can appear in new generations, and the overall transmission spectrum of these quasicrystals shows a fractal nature. As generation number increases, the spectrum reveals self-similarity and characteristic trifurcation of resonances [23].

In what follows, we will consider that layers 'A' and 'B' have the same optical thickness, $n_A d_A = n_B d_B$, and we will introduce the reference wavenumber $\Omega_0$, defined as $n_A d_A \Omega_0 = \pi/2$, which corresponds to the quarter wavelength condition. The ratio $\Omega/\Omega_0$ will be called 'normalized frequency' and will make the comparison of different structures easier. For example, a normalized frequency $\Omega/\Omega_0 = 1$ corresponds to the middle of bandgap region for Bragg crystals, but it is the centre of a wide 'pseudo-pass band' for ThM quasicrystals, a region made of densely located shallow resonances, not always of perfect transmission (figure 2). This band exists only for the specific case of equal optical thicknesses of the layers, and it starts to appear from the second generation, for which formulas (8)-(9) are not valid. The above condition makes them applicable for $n \geq 0$. Moreover, as long as the condition is fulfilled, the transmission spectra of all 1D structures will be translationally invariant along frequency axis, $T(\Omega/\Omega_0 + 2) = T(\Omega/\Omega_0)$ and symmetric with respect to $\Omega/\Omega_0 = 1$.

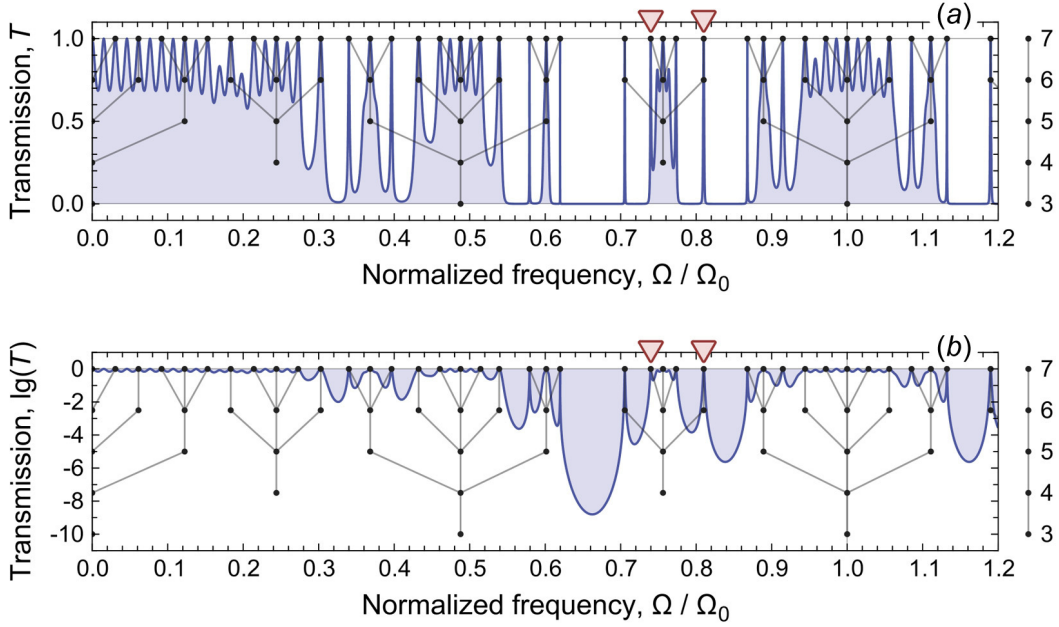

**Figure 2.** (*a*) Linear transmission spectra as a function of normalized frequency ($\Omega/\Omega_0$) for ThM$_7$ quasicrystal with the same set of parameters as in figure 1b. Overlapped diagrams show trifurcation of resonances as generation number increases from 3 to 7 (see vertical axes on the right). (*b*) Same as (*a*), but where band gap regions and their relative strength are emphasized in logarithmic scale.

## 3. Bistability and multistability in Thue-Morse multilayers

The bistability and multistability properties of Bragg gratings have been studied in depth by a number of authors [24-26]. The knowledge of the field profiles is crucial for the proper understanding of bistability and multistability in ThM quasicrystals. In correspondence to the resonance frequencies these field profiles show a self-similar pattern which strongly resembles that of a ThM sequence of smaller generation number (figure 1b). Due to this fact, these resonant states are also called lattice-like states [27]. The level of self-similarity depends on the generation number in correspondence to which a particular resonance first appears. For example, the field profile shown in figure 1b belongs to a ThM quasicrystal of 7th generation. It corresponds to the resonance with normalized frequency $\Omega/\Omega_0 = 0.756$, and, according to the trifurcation diagram shown in figure 2, this resonance is inherited from the 4th generation. Therefore, the field profile mentioned above has the level of similarity equal to $7 - 4 = 3$, and it is possible to distinguish $2^3 = 8$ independent blocks. Moreover, these blocks are arranged according to the ThM sequence of 3rd generation. We have found that for any ThM structure with layers of equal optical thickness, the level of self-similarity is maximal for normalized frequencies belonging to the following series:

$$\Omega/\Omega_0 = \{0,\ 1-\sigma,\ 1,\ 1+\sigma,\ 2,\ 3-\sigma,\ 3,\ 3+\sigma,\ ...\}, \qquad (15)$$

where

$$\sigma = 1 - \frac{2}{\pi}\arctan\left(\sqrt{\frac{2}{(n_A/n_B)+(n_B/n_A)}}\right) \approx 0.512. \qquad (16)$$

However, the localization strength is relatively small at these frequencies, and thus the nonlinear response cannot be enhanced significantly, because the electric field profile is almost flat along the structure. For this purpose, resonances located near the edges of pseudo band gaps are more suitable. In general, the localization strength is inversely proportional to the level of self-similarity, and when the length of the structure increases, the resonant states gradually change from localized to extended ones [28].

The case of Kerr (or cubic) nonlinearity was considered in this work, so that a nonlinear refractive index was additionally taken into account for each type of layer. This leads to an intensity-dependent self-phase modulation, which is able to shift resonance frequencies [25]. The direction of this shift is determined by the sign of the Kerr coefficient. The most interesting cases evidently occur when the sign of the nonlinearity is such that resonances in the spectrum of transmission bend towards the bandgap regions (figures 3-4). In the following, we always choose positive nonlinear coefficients for the two materials, $n_{2A} = 2.5 \cdot 10^{-5}$ cm$^2$/MW and $n_{2B} = 1.0 \cdot 10^{-8}$ cm$^2$/MW (see also figures 3-4) [29-30].

One of the most serious advantages provided by ThM sequence is that for odd generation numbers the corresponding photonic structures are intrinsically asymmetric, and nonlinearity is capable of making transmission sensitive to the propagation direction [16]. This feature is completely absent in nonlinear Bragg structures, where hysteresis curves are the same regardless of the direction of incidence. Although a similar nonreciprocal behavior can be achieved in the framework of linear optics, the latter typically requires making use of magneto-optical media with externally applied static magnetic fields [31] or chiral media such as cholesteric liquid crystals [32]. Moreover, the polarization state is very important for the correct operation of these linear devices, while the design suggested here is free from such complications.

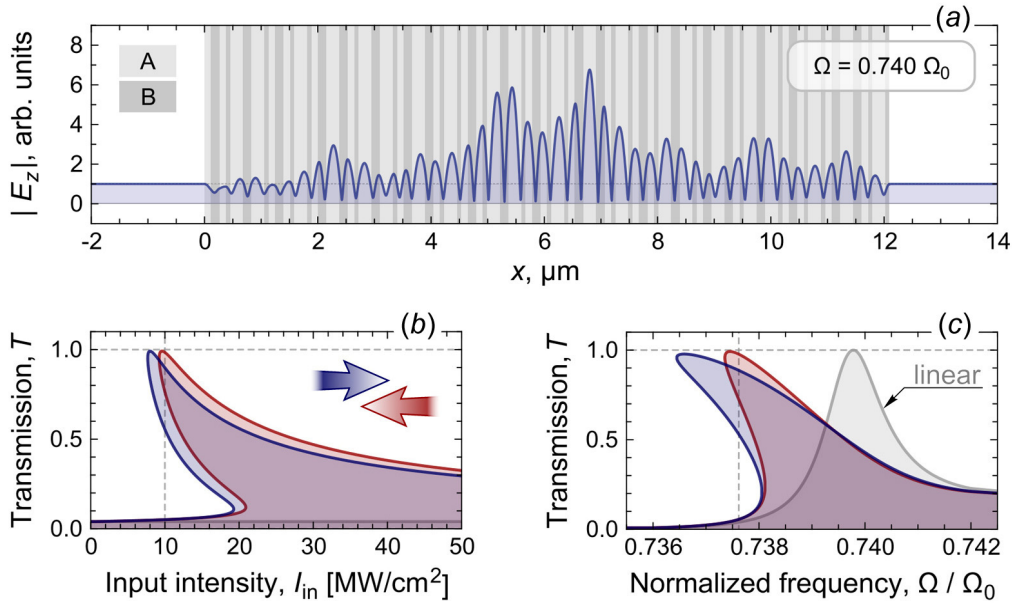

**Figure 3.** (*a*) Electric field profiles at resonance frequency $\Omega/\Omega_0 = 0.740$ for linear ThM$_7$ structure with the same set of parameters as in figure 1b. (*b*) Hysteresis of transmission at $\Omega/\Omega_0 = 0.738$. The nonlinear Kerr coefficients used are $n_{2A} = 2.5 \cdot 10^{-5}$ cm$^2$/MW and $n_{2B} = 1.0 \cdot 10^{-8}$ cm$^2$/MW. The bistability thresholds are different for the forward (blue) and backward (red) incidence. (*c*) Nonlinear transmission spectra for fixed input intensity 10 MW/cm$^2$.

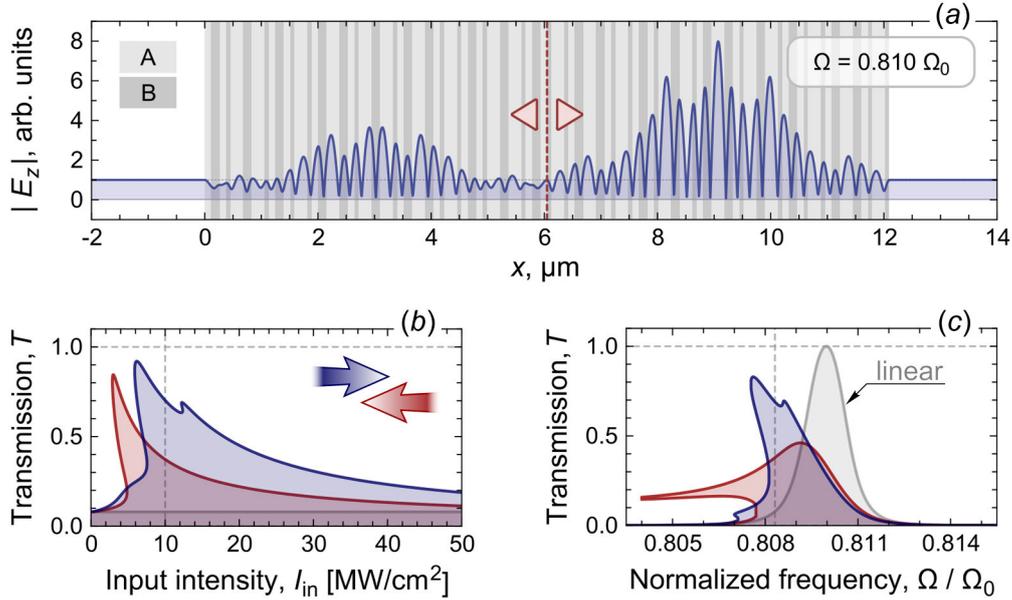

**Figure 4.** (*a*) Electric field profiles at resonance frequency $\Omega/\Omega_0 = 0.810$ for linear ThM$_7$ structure with the same set of parameters as in figure 1b. (*b*) Hysteresis of transmission at $\Omega/\Omega_0 = 0.808$. The nonlinear Kerr coefficients used are $n_{2A} = 2.5 \cdot 10^{-5}$ cm$^2$/MW and $n_{2B} = 1.0 \cdot 10^{-8}$ cm$^2$/MW. The switching thresholds are different for the forward (blue) and backward (red) incidence. (*c*) Nonlinear transmission spectra for fixed input intensity 10 MW/cm$^2$.

The level of self-similarity in the field profiles is also related to the type of hysteresis that can be observed near a particular resonance. In presence of several independent localization centers inside the structure, the interplay between them gives rise to multistability. It is related mainly to the fact that ThM structures of higher generation numbers can be decomposed into those of lower ones. Therefore, to single out a particular hysteresis type in its pure form, it is necessary to check that the corresponding resonance does not occur in previous generations.

Two characteristic examples are shown in figures 3-4. The first case (figure 3) corresponds to a resonance with no self-similarity in the field profile and clearly demonstrates bistability of transmission, whereas in the second case (figure 4) two independent localization centers can be distinguished. The hystereses are shown in figure 3b and 4b, the nonlinear transmission spectra are given in figure 3c and 4c. A detailed explanation of this behavior will be given in the next section. Notice that similarly to Bragg structures, the maxima of electric field can be located mostly in the layers of one type, so that in principle it would be sufficient to use only one nonlinear material.

## 4. Coupled resonator model

### 4.1. Non-separable resonances

In order to provide a qualitative analytical understanding for the nonlinear behavior of ThM quasicrystals, we apply the coupled resonator model, which has been recognized as a useful tool for describing the dynamical properties of single-mode cavities in 2D photonic crystal waveguides [33-35]. Being a phenomenological theory based on general physical concepts like conservation of energy and time-reversal symmetry [36], it is not restricted to the above 2D systems and with little modifications we adapt it to 1D structures. Taking into account the intrinsic spatial asymmetry of resonances, the main equations of this method can be written in the following form

$$\frac{dA}{dt} = -\left[i\omega_0\left(1 - \frac{|A|^2}{P_0}\right) + \gamma\right]A + \kappa_u u_+ + \kappa_v v_+, \quad (17)$$

$$\begin{pmatrix} u_- \\ v_- \end{pmatrix} = \begin{pmatrix} r_u & t \\ t & r_v \end{pmatrix}\begin{pmatrix} u_+ \\ v_+ \end{pmatrix} + A\begin{pmatrix} \kappa_u \\ \kappa_v \end{pmatrix}. \quad (18)$$

In essence, they represent an extension of the scattering matrix method to the time domain. The first equation (17) describes the temporal evolution of a resonant mode in the vicinity of a resonant frequency $\omega_0$. The amplitude of this mode $A$ is normalized in such a way that $|A|^2$ gives the energy inside the cavity. Energy is carried to the cavity from the two ports or channels, which are denoted as $U$ and $V$, located on the left and the right sides, respectively (figure 5a). The amplitudes of the incident signals are denoted as $u_+$ and $v_+$. They are normalized in such a way that $|u_+|^2$ and $|v_+|^2$ give the power flow. The interaction between these signals and the cavity is described by the two coupling coefficients $\kappa_u$ and $\kappa_v$, respectively. Since the cavity not only accumulates the incident energy, but may also partly absorb it or return it back to the channels, there is a damping constant $\gamma$, which characterizes the efficiency of these processes. It can be related to the quality factor of the resonator $Q = \omega_0/(2\gamma)$. As to the influence of nonlinearity, it results in the shift of the resonant frequency. In equation (17), we introduce a characteristic energy $P_0$ that allows to take this shift into account [37], which can be either positive or negative depending on the common sign of the Kerr coefficients in the multilayered structure.

The second equation (18) describes the output from the cavity. The right hand side of this equation consists of two parts. The first part characterises the background reflection from the cavity as if the amplitude of the mode inside it were zero. In fact, it has exactly the form of a scattering matrix. The background transmission coefficient $t$ is equal for both directions even for nonsymmetric structures, while the reflection coefficients in general have different phases and are denoted as $r_u$ and $r_v$. As to the second part of equation (18), it is proportional to the amplitude of the mode, and thus provides the corrections caused by the interaction of incident signals with the cavity.

Assuming that the system operates at a fixed frequency, it is possible to derive from equations (17), (18) an explicit expression for the nonlinear transmission spectrum along the forward (backward) propagation direction:

$$T^{u,v} \equiv \frac{I_{\text{out}}^{u,v}}{I_{\text{in}}^{u,v}} = \frac{\eta}{1 + (\delta + I_{\text{out}}^{u,v}/I_0^{u,v})^2}, \quad (19)$$

where $I_{\text{in}}^u = |u_+|^2$ ($I_{\text{in}}^v = |v_+|^2$) is the input intensity, $I_{\text{out}}^u = |v_-|^2$ ($I_{\text{out}}^v = |u_-|^2$) is the output intensity, $\delta = (\omega - \omega_0)/\gamma$ is the normalized detuning from the resonance frequency, and $\eta = |\kappa_u \kappa_v|^2/\gamma^2$ corresponds to the maximum allowed transmission. The characteristic intensities $I_0^{u,v}$, and this is a crucial observation for the present work, are in general different for the incidence from the left and from the right

$$I_0^{u,v} = (\gamma/\omega_0)P_0|\kappa_{u,v}|^2. \quad (20)$$

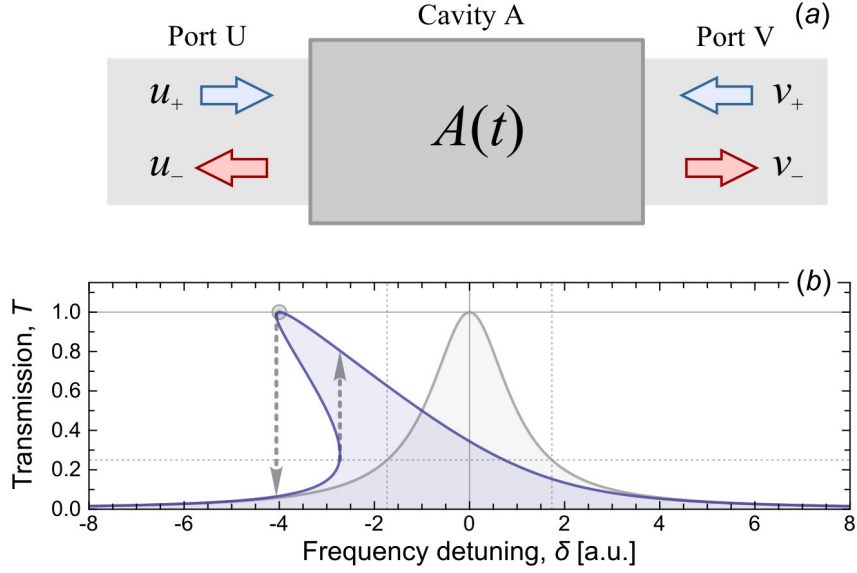

**Figure 5.** (*a*) The scheme describing the quantities used in the equations (17)-(18) for the coupled resonator model. (*b*) The typical form of nonlinear transmission spectrum following from formula (19).

In the case of single Dirac-delta-like nonlinear layer in the system, formula (19) coincides with a previously investigated exact analytical solution [38]. However, taking into account a continuously distributed nonlinearity gives rise to a new feature: if the structure is spatially asymmetric, the characteristic intensity can depend on the direction of incidence, as it is evident from equation (20).

Since the formula (19) is equivalent to a third-order polynomial in the output intensity $I_{\text{out}}^{u,v}$, by definition it fails to describe resonances with multistability, but it gives a good approximation for resonances with a strictly bistable response (compare for instance figure 5b with figure 3b-c).

### 4.2. Separable resonances

In the more complex case, when the field profiles consist of several independent parts, one simply needs to increase the number of equations in the coupled resonator model. There should be exactly one set of equations (17)-(18) for each separable resonance. For example, the case shown in figure 4 can be described by the following system of equations

$$\frac{dA}{dt} = -\left[i\omega_0\left(1 - \frac{|A|^2}{P_a}\right) + \gamma_a\right]A + \kappa_a(u_+ + s_+), \quad (21)$$

$$\frac{dB}{dt} = -\left[i\omega_0\left(1 - \frac{|B|^2}{P_b}\right) + \gamma_b\right]B + \kappa_b(v_+ + s_-), \quad (22)$$

$$\begin{pmatrix} u_- \\ s_- \end{pmatrix} = \exp(i\varphi_a)\begin{pmatrix} u_+ \\ s_+ \end{pmatrix} + A\kappa_a, \quad (23)$$

$$\begin{pmatrix} v_- \\ s_+ \end{pmatrix} = \exp(i\varphi_b)\begin{pmatrix} v_+ \\ s_- \end{pmatrix} + B\kappa_b, \quad (24)$$

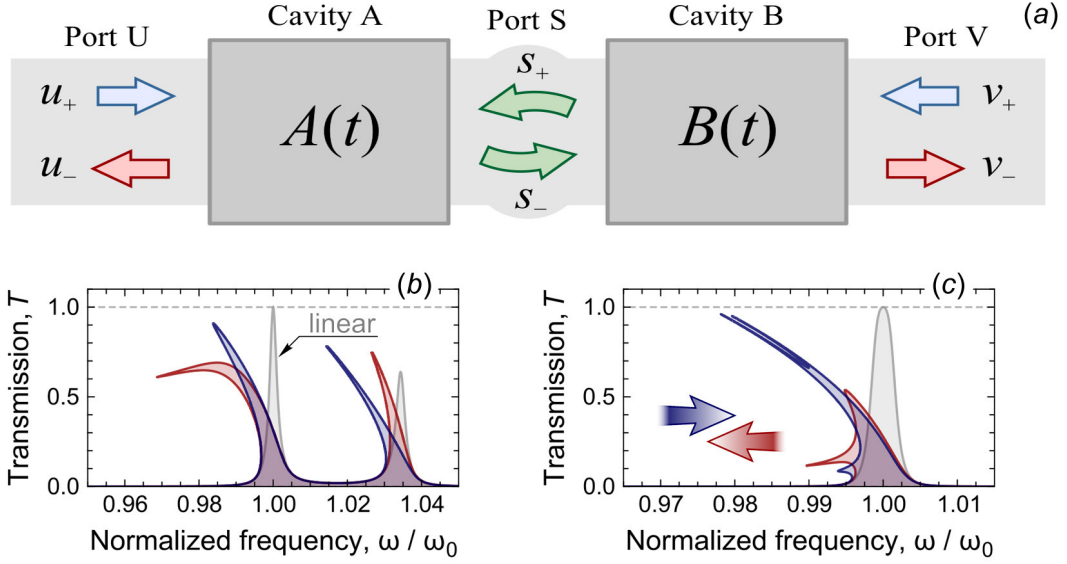

**Figure 6.** (*a*) The scheme describing quantities used in the equations (21)-(24) of the coupled resonator model. (*b*, *c*) The typical form of nonlinear transmission spectra following from the formula (25), when the dimensionless parameters are set to $\omega_0 = 800$, $\gamma_a = 2.89$, $\gamma_b = 1.44$, $P_a = 120$, $P_b = 80$. The only difference between figures (*b*) and (*c*) is in the phase mismatch $\varphi_a + \varphi_b = 0.05\pi$ and $\varphi_a + \varphi_b = \pi$, respectively. The amplitudes of incident signals are set to $u_+ = 2$ for the propagation in forward direction (blue lines) and $v_+ = 2$ for the backward direction (red lines). The linear transmission spectra are given by gray lines.

where $A$ and $B$ give the amplitudes of the two separable resonances, which interact through a virtual port $S$ (figure 6a). These resonances have the same resonance frequency $\omega_0$, but in general their damping constants $\gamma_{a,b}$ and characteristic energies $P_{a,b}$ will be different. Since both separable resonances shown in figure 4 are symmetric, the coupling to either of them can be described by single coefficient $\kappa_{a,b} = \sqrt{\gamma_{a,b}}$. Moreover, because the common resonant frequency is located in the bandgap (or pseudo band gap) region, the background transmission can be neglected, and the scattering matrix entering equation (18) can be replaced by a reflection coefficient, which is equal for both directions of incidence and has the form $\exp(i\varphi_{a,b})$.

The system (21)-(24) can be readily solved in the frequency domain. We found the following formula, which describes the nonlinear transmission spectra with multistable behaviour:

$$T = \frac{1}{|\xi_a\xi_b + (1-i\xi_a)(1-i\xi_b)\exp[-i(\varphi_a + \varphi_b)]|^2}. \qquad (25)$$

Expression (25) contains two parameters, $\xi_a$ and $\xi_b$, which can be calculated given the output signal $v_-$ (or $u_-$) by using the following ladder:

$$B = v_- / \kappa_b, \qquad (26)$$

$$\xi_b = \delta_b + \frac{\omega_0 |B|^2}{\gamma_b P_b}, \qquad (27)$$

$$A = (1 - i\xi_b + i\xi_b \exp[i(\varphi_a + \varphi_b)])(v_- / \kappa_a), \tag{28}$$

$$\xi_a = \delta_a + \frac{\omega_0 |A|^2}{\gamma_a P_a}. \tag{29}$$

The typical nonlinear transmission spectra calculated by using formula (25) are shown in figure 6b-c. The main parameter which determines the shape of nonlinear transmission spectra is the phase mismatch $\varphi_a + \varphi_b$. It describes the efficiency of interaction between separable resonances and is responsible for the splitting of initially identical resonant frequencies even in linear case. The only difference between figures 6b and 6c is in the phase mismatch, and the two well-separated resonances shown in the first case become undistinguishable for small intensities in the second case (see gray lines). As the intensity of incident signal increases, this degeneracy disappears giving rise to multistability (see red and blue lines for the forward and backward incidence, respectively). Qualitatively, the nonlinear transmission spectra shown in figure 6c corresponds to that in figure 4c.

It is worth noting that if $\varphi_a + \varphi_b = 0$, the formula (25) reduces to

$$T = \frac{1}{1 + (\xi_a + \xi_b)^2}, \tag{30}$$

which means that the two resonances do not interact with each other and the resulting nonlinear transmission spectra reveals only a bistable behaviour.

## 5. Nonlinear optical diode action in Thue-Morse multilayers

The difference in bistability thresholds can be small, but it creates favorable conditions for unidirectional propagation, so that ThM quasicrystals can be used as all-optical diodes [16]. Similar to electronic circuits, these devices are indispensable, when there is a need to suppress the flow of light in one direction or to avoid problems caused by unwanted reflections. Various types of optical diodes have been proposed and realized, which can work both in linear [31-32] and nonlinear regime [39-42]. The primary figure of merit, which determines the efficiency of this device, is the contrast ratio $C = T_f / T_b$ between transmission along the forward ($T_f$) and the backward ($T_b$) directions. $C$ can be very large in ThM structures due to the fact that this device can operate on two different hysteresis branches depending on propagation direction (figure 3b).

It is possible to use the intensities of both up and down transitions to achieve strongly nonreciprocal transmission. In the former case, the scheme works in passive mode, but the maximum value of transmission is limited [43]. In the latter case, the transmission can be almost perfect, but the scheme requires an additional short pump signal in order to switch to the higher stable branch of hysteresis [44].

The FDTD method was applied to demonstrate the switching dynamics in time domain (figure 7). To maintain the second order accuracy of the Yee-scheme in case of multilayered structures, nonuniform spatial mesh was used with electric field nodes aligned to the boundaries between layers (see Appendix). Since the structure can accumulate and release energy, the sum of reflected (given by red dotted line in figure 7) and transmitted intensities (blue dashed line) should not necessarily give the incident one (gray solid line). This constraint becomes valid only after all the transient dynamics is over.

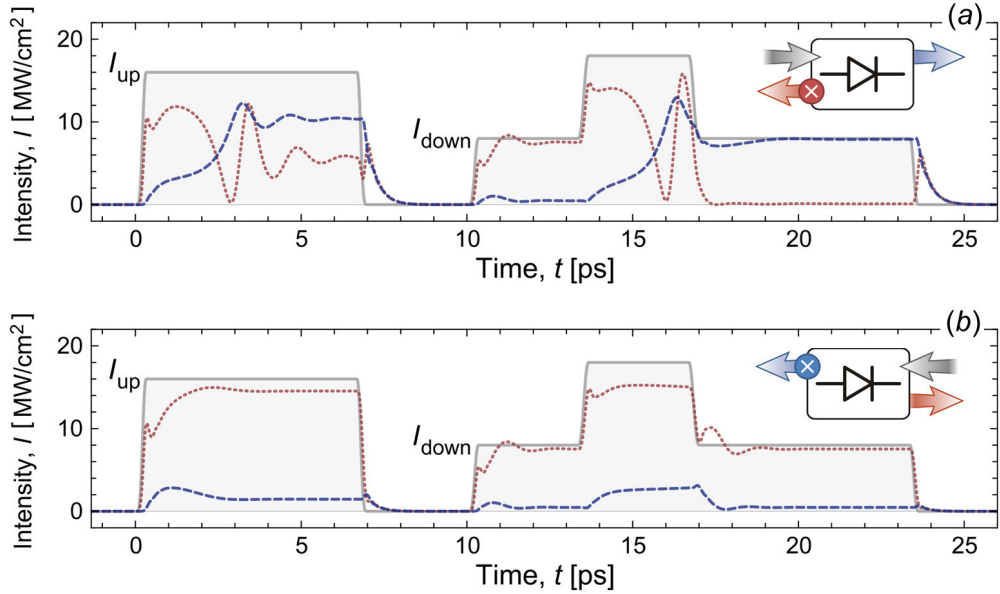

**Figure 7.** Numerical simulations of nonlinear ThM$_7$ structure acting as an optical diode for the forward (*a*) and backward (*b*) incidence at the frequency 0.738 $\Omega_0$. Left parts of the figures refer to the passive scheme and right ones to the active scheme when an additional short pump signal is applied to facilitate switching. The intensities of incident (solid), reflected (dotted), and transmitted (dashed) signals are given as a function of time.

The input intensity was set first to the value, which is sufficiently large for the up switching in the forward direction, but at the same time is smaller than corresponding value for the backward direction. The maximum transmission obtained was $T_f = 65\%$ with the contrast ratio $C = 7.2$, which is quite close to the theoretical limit of $C_{max} = 9$ following from (31) in the assumption of passive scheme [43]. In the second stage, the input intensity was set just above the down switching threshold for the forward direction, which ensures transmission of almost $T_f = 100\%$ in this direction, while in the opposite direction transmission will be strongly suppressed due to the presence of the pseudo band gap. The contrast ratio achieved was $C = 20$, but it is not limited in principle. The major drawback of this scheme is that it is necessary to force the system to switch to the upper branch of hysteresis whether by applying an auxiliary pump signal, or by temporarily increasing the input intensity.

## 6. Conclusions and future work

The nonlinear properties of quasicrystals based on ThM sequence were investigated. It was shown that the interplay between the intrinsic spatial asymmetry of these structures for odd generation numbers and Kerr nonlinearity makes the switching thresholds induced by bistability and multistability sensitive to the propagation direction. The role of self-similarity was emphasized to explain the shape of hysteresis curves observed near a particular resonance, and conditions necessary to achieve highly non-reciprocal propagation were formulated. Coupled resonator model was used as a phenomenological model. FDTD simulations have been used to confirm the results obtained by the nonlinear transfer matrix method, and to show how ThM multilayers can exhibit an effective optical diode action, both in passive and active operation.

**Acknowledgments**
This work was supported by the German Max Planck Society for the Advancement of Science (MPG).

**Appendix**

Maxwell's equations in differential form

$$\frac{\partial E_z(x,t)}{\partial x} = \frac{1}{c}\frac{\partial H_y(x,t)}{\partial t}, \tag{A.1}$$

$$\frac{\partial H_y(x,t)}{\partial x} = \frac{1}{c}\frac{\partial D_z(x,t)}{\partial t}, \tag{A.2}$$

can be approximated by finite differences

$$H_y^{n+1/2}(i+1/2) = H_y^{n-1/2}(i+1/2) + \frac{c\Delta t}{\Delta x}\left[E_z^n(i+1) - E_z^n(i)\right], \tag{A.3}$$

$$D_z^{n+1}(i) = D_z^n(i) + \frac{c\Delta t}{\Delta x}\left[H_y^{n+1/2}(i+1/2) - H_y^{n+1/2}(i-1/2)\right], \tag{A.4}$$

where integer indices $n$ and $i$ serve as temporal and spatial coordinates for the Yee-scheme [45]. Courant stability condition requires $c\Delta t \leq \Delta x$, and it will be assumed that $c\Delta t = \Delta x/2$. The spatial grid should be fine enough to capture the smallest wavelength which can be excited in the system. It is usually chosen as $\Delta x = \lambda_{\min}/10$ for uniform grids. However, nonuniform grids can be more advantageous for simulating multilayered structures.

To tailor two different uniform grids, it is convenient to use the integral representation of Maxwell's equations

$$E_z(x_L,t) - E_z(x_R,t) = -\frac{1}{c}\frac{\partial}{\partial t}\int_{x_L}^{x_R} H_y(x,t)dx, \tag{A.5}$$

$$-H_y(x_L,t) + H_y(x_R,t) = \frac{1}{c}\frac{\partial}{\partial t}\int_{x_L}^{x_R} D_z(x,t)dx. \tag{A.6}$$

After integration over the boundary between two layers, the following finite difference equations can be obtained (the symbol used are defined in figure A1)

$$H_y^{n+1/2}(\beta_b) = H_y^{n-1/2}(\beta_b) + \frac{c\Delta t}{\rho_a + \rho_b}[E_z^n(\rho_b) - E_z^n(-\rho_a)], \tag{A.7}$$

$$E_z^{n+1}(\rho_b) = E_z^n(-\rho_a) + \frac{c\Delta t}{\varepsilon_a\beta_a + \varepsilon_b\beta_b}\left[H_y^{n+1/2}(\beta_b) - H_y^{n+1/2}(-\beta_a)\right]. \tag{A.8}$$

It can be proven that these equations are valid up to the second order, if the nodes of electric field are placed on the boundaries between layers, and the steps of the spatial grids inside layers of different types satisfy $\varepsilon_a(\Delta x_a)^2 = \varepsilon_b(\Delta x_b)^2 = (\Delta x)^2$. This condition can be exactly fulfilled if layers composing the structure are of equal optical thickness. In general case, the second-order error of the scheme with nonuniform mesh is estimated to be

$$\varepsilon_b(\Delta x_b/m_b)\delta, \tag{A.9}$$

where $m$ is the number of nodes inside an unmatched layer, and $\delta$ is a missing or superfluous thickness that caused the mismatch. When the scheme with uniform mesh is used, the error is proportional to the difference in permittivities and cannot be turned into zero

$$(\varepsilon_b - \varepsilon_a)(\Delta x/2)^2. \tag{A.10}$$

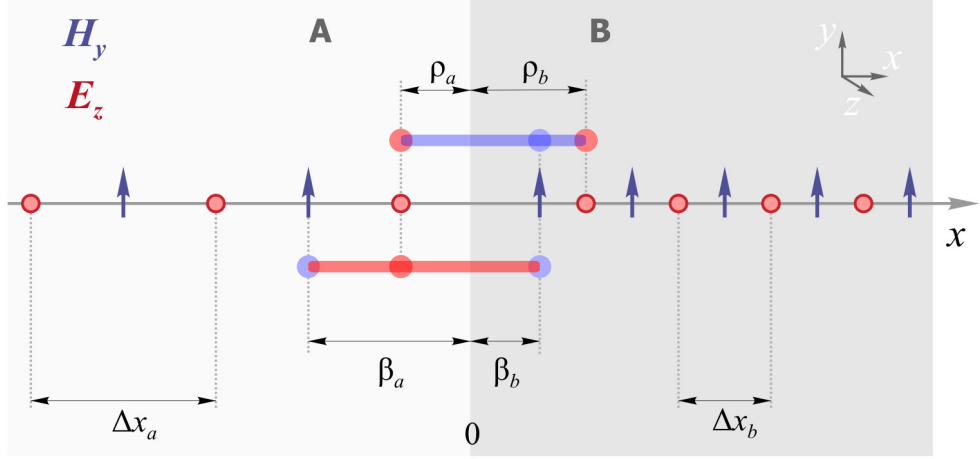

**Figure A1.** The scheme describing quantities used in the tailoring of two uniform grids at the boundary between dielectric media.

The constituent relation with instantaneous Kerr-nonlinearity is considered

$$D_z = \varepsilon_L E_z + \varepsilon_K E_z^3. \qquad (A.11)$$

According to (A.6) and (A.8), the linear $\varepsilon_L$ and nonlinear $\varepsilon_K$ contributions to permittivity should be averaged at the boundaries between layers

$$\varepsilon_{L,\text{eff}}(i) = (\varepsilon_{L,a}\Delta x_a + \varepsilon_{L,b}\Delta x_b)/\Delta x, \qquad (A.12)$$

$$\varepsilon_{K,\text{eff}}(i) = (\varepsilon_{K,a}\Delta x_a + \varepsilon_{K,b}\Delta x_b)/\Delta x. \qquad (A.13)$$

The equation (A.11) can be written in finite differences as [46]

$$E_n = [D_n + 2\varepsilon_K (E_{n-1})^3]/[\varepsilon_L + 3\varepsilon_K (E_{n-1})^2], \qquad (A.14)$$

which is equivalent to the solution of the cubic equation (A.11) by iteration method and is always stable for positive $\varepsilon_L$ and $\varepsilon_K$. In contrast to equation

$$D_z = \varepsilon_L E_z + (3/4)\varepsilon_K |E_z|^2 E_z, \qquad (A.15)$$

the usage of constituent relation (A.11) allows taking into account the generation of third harmonic, but attention should be paid to the correct definition of $\varepsilon_K$.

To separate incident and reflected waves from the structure, the so-called Total-Field Scattered-Field technique is used on the left (and of the right) of the structure

$$H_y^{n+1/2}(i_F \mp 1/2) = H_y^{n-1/2}(i_F \mp 1/2) \pm \frac{c\Delta t}{\Delta x}\left[E_z^n(i_F) - E_{z,\text{inc}}^n(i_F) - E_z^n(i_F \mp 1)\right], \qquad (A.16)$$

$$D_z^{n+1}(i_F) = D_z^n(i_F) + \frac{c\Delta t}{\Delta x}\left[H_y^{n+1/2}(i_F + 1/2) - H_y^{n+1/2}(i_F - 1/2) \mp H_{y,\text{inc}}^{n+1/2}(i_F \mp 1/2)\right]. \qquad (A.17)$$

Sources of electromagnetic field on the left (and on the right) are introduced as

$$E_{z,\text{inc}}^n(i_S) = E_{z,\text{inc}}(n\,c\Delta t), \qquad (A.18)$$

$$H_{y,\text{inc}}^{n+1/2}(i_S \mp 1/2) = \mp E_{z,\text{inc}}((n+3/2)c\Delta t). \qquad (A.19)$$

The Absorbing Boundary Conditions in case when $c\Delta t = \Delta x/2$ take the following form on the left (and on the right) boundary of the simulation region

$$H_y^{n+1/2}(i_B \mp 1/2) = H_y^{n-3/2}(i_B \pm 1/2). \quad (A.20)$$

Smooth envelops for switching on and off monochromatic signals are defined by splines of 6th order, which ensure the continuity of fields up to the second derivative [47]

$$U(t) = \begin{cases} u(t/\tau) & \text{for } 0 < t < \tau, \\ 1 & \text{for } \tau \leq t \leq T - \tau, \\ u[(T-t)/\tau] & \text{for } T - \tau < t < T, \end{cases} \quad (A.21)$$

where $T$ is the total time when the signal is nonzero, $\tau$ is the duration of the switching, and $u(x) = 10x^3 - 15x^4 + 6x^5$. The intensities of reflected and transmitted waves $A$ can be extracted from the simulation through the following integration

$$(1/P)\int_t^{t+P} A\cos(m\omega t - \varphi)\exp(in\omega t)dt = (A/2)\exp(i\varphi)\delta_{nm}, \quad (A.22)$$

where $P = 2\pi/\omega$ is the period of carrying wave.